# Hybrid Stars in the Framework of the Local Nambu-Jona-Lasinio Model for Quark Matter

G. B. Alaverdyan[*)] and Yu. L. Vartanyan

*Yerevan State University, A.Manoogyan str. 1, 0025 Yerevan, Armenia*

*The integral parameters of neutron stars are studied taking into account the hadron-quark phase transition, which leads to the formation of a core of quark matter in the central part of a star. The quark matter is described using the local Nambu-Jona-Lasinio (NJL) model. The thermodynamic characteristics of the hadron matter are calculated in the framework of an extended version of the relativistic mean field (RMF) model that includes the contribution of the scalar-isovector $\delta$-meson effective field. A Maxwell construction is used to determine the parameters of the phase transition. It is shown that for the equation of state examined here, stable hybrid stars correspond to a narrow range of values for the central density $\rho_c \in (1.71 \div 1.73] \cdot 10^{15} \, g/cm^3$. In our model hybrid stars lie on the same branch as neutron stars, so that a branch with a third family is not formed. The maximum mass of a stable hybrid star is found to be $M_{\max} = 2.05 \, M_\odot$. The configuration with the maximum mass has a quark core with mass $M_{core} \approx 10^{-3} \, M_\odot$ and radius $R_{core} \approx 0.6 \, km$.*

---

[*)] e-mail: galaverdyan@ysu.am

# 1. Introduction

Studies of the hadron-quark phase transition at extremely high densities is one of the most important areas in modern physics. Many papers in high energy and elementary particle physics, as well as in the physics of compact cosmic objects (neutron and quark stars) have been devoted to clarifying the possible consequences of this transition and gathering information on properties of the quark matter formed by quark deconfinement.

It is known that the properties of compact stars depend on the equation of state of matter over a rather wide range of densities. The lower bound of this range corresponds to the density of ordinary matter with an atomic and molecular composition. The upper bound may exceed the densities of nucleon matter in atomic nuclei by an order of magnitude or more. At these extremely high densities in baryon matter, when quarks are "trapped" inside hadrons, a deconfinement phase transition can take place, leading to the formation of quark matter.

In order to obtain the equation of state for neutron star matter including a possible hadron-quark phase transition, it is necessary to have the equations of state for both hadron and quark matter, separately, but knowledge of the type of phase transition is also required. In fact, depending on the coefficient of surface tension between the quark and hadron matter, the phase transition can proceed according to two different scenarios. When this coefficient is comparatively low, the phase transition involves formation of an electrically neutral hadron-quark mixed phase [1]. Then the surface tension coefficient is large, each phase will be electrically neutral separately and the phase transition will take place at a constant pressure determined by the Maxwell construction. In the latter case, the density undergoes a discontinuity during the transition. The currently available information on the coefficient of surface tension between hadron and quark matter is not sufficient to provide a unique determination of which phase transition scenario will take place. In this paper we assume that the transition follows the Maxwell scenario.

The lack of a unified strict theory for an adequate explanation of the properties of matter over such a wide range of densities makes it necessary to use different model approaches which are valid only within one or another limited region. We have used the relativistic mean field (RMF) theory [2-5] to describe the properties of hadron matter in the region of a possible first order phase transition.

Besides the well-known phenomenological MIT quark bag model [6], the Nambu-Jona-Lasino (NJL) model [7,8], which successfully reproduces many features of quantum chromodynamics [9-11], is currently used to study the possible existence of quark matter in the central part of neutron stars.

In our earlier papers based on the MIT bag model we have studied hybrid stars [12-15] and strange stars [16-19].

This paper is a study of the thermodynamic properties of quark matter in terms of a local version of the NJL model. The possible existence of quark matter in the interior of neutron stars is clarified and we test the consistency of the results with the limit $M_{\max} > 2 M_\odot$ on the maximum mass of hybrid stars that follows from the observed existence of pulsars with masses on the order of two solar masses [20,21].

## 2. The quark phase

Strongly interacting three-flavor quark matter at zero temperature was modelled using a local NJL model, which is an effective theory of quantum chromodynamics (QCD) that also accounts for nonperturbative effects such as the partial recovery of chiral symmetry at high densities. In this paper the following expression [22]* is used for the lagrangian:

$$\mathcal{L}_{NJL} = \overline{\psi}\left(i\gamma^\mu \partial_\mu - \hat{m}_0\right)\psi + G \sum_{a=0}^{8}\left[\left(\overline{\psi}\lambda_a\psi\right)^2 + \left(\overline{\psi} i\gamma_5 \lambda_a \psi\right)^2\right] - \\ - K\left\{\det{}_f\left(\overline{\psi}(1+\gamma_5)\psi\right) + \det{}_f\left(\overline{\psi}(1-\gamma_5)\psi\right)\right\}, \quad (1)$$

where $\psi$ is the fermion quark spinor field $\psi_f^c$ with three flavors $f$ = 1, 2, 3 and three colors $c$ = $r$, $g$, $b$, $\hat{m}_0$ is the diagonal mass matrix of the current quarks in the flavor space $\hat{m}_0 = \text{diag}(m_{0u}, m_{0d}, m_{0s})$, $\lambda_a$ ($a$ = 1, 2, ..., 8) are the Gell-Mann SU(3) matrices in flavor space, $\lambda_0 = \sqrt{2/3}\, I$, $G$ is the constant for the scalar channel of a four-quark interaction, and $K$ is the constant for a Kobayashi-Maskawa-'t Hooft six-quark interaction [23].

The characteristics of the quark matter in the mean field approximation are expressed in terms of the quark chiral condensates

$$\sigma_u = \langle\overline{\psi}_u \psi_u\rangle, \quad \sigma_d = \langle\overline{\psi}_d \psi_d\rangle, \quad \sigma_s = \langle\overline{\psi}_s \psi_s\rangle.$$

The quarks are treated as quasiparticles with constituent masses $m_i$ ($i = u, d, s$) that satisfy the gap equations

$$\begin{aligned} m_u &= m_{0u} - 4G\sigma_u + 2K\sigma_d \sigma_s, \\ m_d &= m_{0d} - 4G\sigma_d + 2K\sigma_s \sigma_u, \\ m_s &= m_{0s} - 4G\sigma_s + 2K\sigma_u \sigma_d. \end{aligned} \quad (2)$$

The quark condensates $\sigma_i$ are, in turn, given in terms of the constituent masses by

$$\sigma_i = -\frac{3}{\pi^2} \int_{p_F(n_i)}^{\Lambda} \frac{m_i}{\sqrt{k^2 + m_i^2}} k^2 dk \quad (i = u, d, s), \quad (3)$$

where $n_i$ is the concentration, $p_F(n_i) = \left(\pi^2 n_i\right)^{1/3}$ is the fermi-momentum of a quark with flavor $i$, and $\Lambda$ is the momentum at the ultraviolet cutoff, which will be needed in connection with the non-renormalizability of the NJL

---

* We use the natural system of units $\hbar = c = 1$.

model.

The gap equations (2) imply that a contribution to the mass of a quark with given flavor comes not only from the quark condensate of this flavor, but also from the condensates of the other two flavors. This mixing of flavors in the masses of the constituent quarks happens because of the presence, in the lagrangian, of a term for the six-quark 'tHooft interaction.

The condition of local electrical neutrality for a plasma consisting of *u*, *d*, and *s*-quarks and electrons has the form

$$\frac{2}{3}n_u - \frac{1}{3}n_d - \frac{1}{3}n_s - n_e = 0. \qquad (4)$$

The baryon concentration is given in terms of the concentrations of the quarks by

$$n_B = \frac{1}{3}(n_u + n_d + n_s). \qquad (5)$$

The $\beta$-equilibrium condition leads to relationships between the chemical potentials of the quarks and electrons:

$$\mu_d(n_d, m_d) = \mu_u(n_u, m_u) + \mu_e(n_e), \qquad (6)$$

$$\mu_s(n_s, m_s) = \mu_d(n_d, m_d). \qquad (7)$$

The chemical potentials of the quarks and electrons are given by

$$\mu_i(n_i, m_i) = \sqrt{(\pi^2 n_i)^{2/3} + m_i^2} \quad (i = u, d, s),$$
$$\mu_e(n_e) = \sqrt{(3\pi^2 n_e)^{2/3} + m_e^2}. \qquad (8)$$

Solving the system (2)-(7) of ten equations with a specified value of the baryon concentration $n_B$ yields the constituent masses of the quarks $m_u$, $m_d$, and $m_s$, the concentrations of the particles $n_u$, $n_d$, $n_s$, and $n_e$, and the values of the quark condensates $\sigma_u, \sigma_d$, and $\sigma_s$.

The energy density and pressure of an electrically neutral plasma consisting of *u*, *d*, and *s*-quarks and electrons and described by the lagrangian (1) in the NJL model are given by

$$\varepsilon_{udse}(n_B) = \frac{3}{\pi^2} \sum_{i=u,d,s} \int_0^{(\pi^2 n_i(n_B))^{1/3}} \sqrt{k^2 + m_i(n_B)^2} \, k^2 dk +$$
$$+ 2G\left[\sigma_u(n_B)^2 + \sigma_d(n_B)^2 + \sigma_s(n_B)^2\right] -$$
$$- 4K\sigma_u(n_B)\sigma_d(n_B)\sigma_s(n_B) + \frac{1}{\pi^2} \int_0^{(3\pi^2 n_e(n_B))^{1/3}} \sqrt{k^2 + m_e^2} \, k^2 dk + \qquad (9)$$
$$+ \frac{3}{\pi^2} \sum_{i=u,d,s} \int_0^{\Lambda} \left(\sqrt{k^2 + m_i(0)^2} - \sqrt{k^2 + m_i(n_B)^2}\right) k^2 dk -$$
$$- 2G\left[\sigma_u(0)^2 + \sigma_d(0)^2 + \sigma_s(0)^2\right] + 4K\sigma_u(0)\sigma_d(0)\sigma_s(0),$$

$$P_{udse}(n_B) = \sum_{i=u,d,s} n_i(n_B)\sqrt{(\pi^2 n_i(n_B))^{2/3} + m_i(n_B)^2} +$$
$$+ n_e(n_B)\sqrt{(3\pi^2 n_e(n_B))^{2/3} + m_e^2} - \varepsilon_{udse}(n_B). \qquad (10)$$

The parameters of the model are the quark current masses $m_{0u}, m_{0d}$, and $m_{0s}$, the cutoff parameter $\Lambda$, and the coupling constants $G$ and $K$. In Ref. 22 these parameters were determined under the assumption of isospin symmetry, i.e., $m_{0u} = m_{0d}$. A set of these parameters was found that best reproduces the experimental values of the masses of the pseudoscalar mesons $\pi$, $K$, and $\eta'$, along with that of the pion decay constant $f_\pi$. In this paper the numerical calculations were done using the values of the model parameters given in Ref. 22:

$$m_{0u} = m_{0d} = 5.5 \text{ MeV}, \; m_{0s} = 140.7 \text{ MeV}, \; \Lambda = 602.3 \text{ MeV},$$
$$G = 1.835/\Lambda^2, \; K = 12.36/\Lambda^5.$$

Equations (9) and (10) yield a parametric form of the equation of state for quark matter consisting of free $u$, $d$, and $s$-quarks, and electrons. Note that these expressions are valid for baryon concentrations that satisfy the condition

$$\mu_d(n_B) \geq m_s(n_B). \qquad (11)$$

The threshold for formation of an $s$-quark is determined by the equation $\mu_d(n_B^{cr}) = m_s(n_B^{cr})$. Below this threshold the quark matter consists of $u$ and $d$ quarks and electrons. The parameters of the quark matter for a specified baryon concentration at densities below the threshold for formation of a strange quark ($n_B < n_B^{cr}$) are determined by solving the nine equations (2)-(6) with the additional condition $n_s = 0$. Denoting these solutions by $\tilde{m}_u, \tilde{m}_d, \tilde{m}_s, \tilde{n}_u, \tilde{n}_d, \tilde{n}_e, \tilde{\sigma}_u, \tilde{\sigma}_d, \tilde{\sigma}_s$, for the equation of state of quark matter consisting of $u$ and $d$-quarks and electrons we have

$$\varepsilon_{ude}(n_B) = \frac{3}{\pi^2} \sum_{i=u,d} \int_0^{(\pi^2 n_i(n_B))^{1/3}} \sqrt{k^2 + \tilde{m}_i(n_B)^2}\, k^2 dk + \\
+ 2G\left[\tilde{\sigma}_u(n_B)^2 + \tilde{\sigma}_d(n_B)^2 + \tilde{\sigma}_s(n_B)^2\right] - \\
- 4K\tilde{\sigma}_u(n_B)\tilde{\sigma}_d(n_B)\tilde{\sigma}_s(n_B) + \frac{1}{\pi^2}\int_0^{(3\pi^2 \tilde{n}_e(n_B))^{1/3}} \sqrt{k^2 + m_e^2}\, k^2 dk + \\
+ \frac{3}{\pi^2}\sum_{i=u,d,s}\int_0^{\Lambda}\left(\sqrt{k^2 + \tilde{m}_i(0)^2} - \sqrt{k^2 + \tilde{m}_i(n_B)^2}\right)k^2 dk - \\
- 2G\left[\tilde{\sigma}_u(0)^2 + \tilde{\sigma}_d(0)^2 + \tilde{\sigma}_s(0)^2\right] + 4K\tilde{\sigma}_u(0)\tilde{\sigma}_d(0)\tilde{\sigma}_s(0), \quad (12)$$

$$P_{ude}(n_B) = \sum_{i=u,d} \tilde{n}_i(n_B)\sqrt{(\pi^2 \tilde{n}_i(n_B))^{2/3} + \tilde{m}_i(n_B)^2} + \\
+ \tilde{n}_e(n_B)\sqrt{(3\pi^2 \tilde{n}_e(n_B))^{2/3} + m_e^2} - \varepsilon_{ude}(n_B). \quad (13)$$

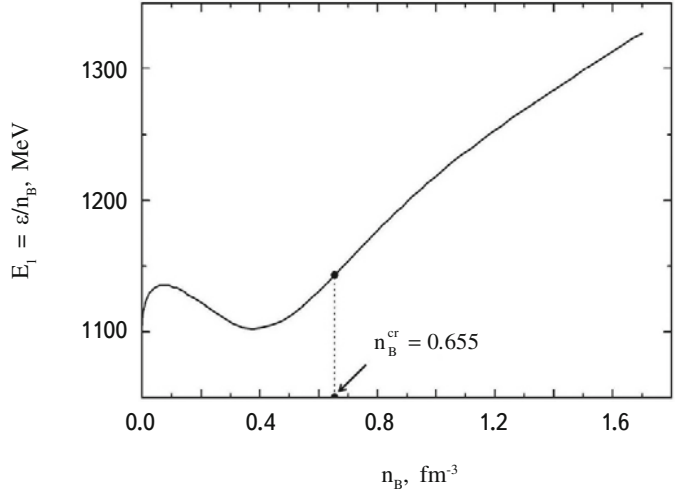

Fig. 1. The energy of quark matter per baryon as a function of baryon concentration. The point indicated on the curve corresponds to the threshold for creation of a strange quark.

Equations (12) and (13) determine the parametric form of the equation of state for quark matter made up of free $u$ and $d$-quarks and electrons. The baryon chemical potential plays an important role in finding the parameters of the chemical equilibrium between the two phases. For the baryon chemical potential of quark matter we have

$$\mu_B^{QM}\left(P^{QM}\right) = \frac{P^{QM} + \varepsilon^{QM}\left(P^{QM}\right)}{\varepsilon_B^{QM}\left(P^{QM}\right)}. \tag{14}$$

Figure 1 shows the results of a numerical calculation of the energy per baryon, $E_1 = \varepsilon/n_B$, as a function of baryon concentration $n_B$. The point indicated on the curve corresponds to the threshold $n_B^{cr} = 0.655$ fm$^{-3}$ for creation of strange quarks. This figure shows that the energy per baryon exceeds the corresponding value for the more coupled nucleus $M_{^{56}Fe}/56 = 930.4$ MeV. This means that Witten's hypothesis that the ground state of matter at zero pressure is a β-equilibrium of three-flavor $uds$ matter is not realized in the NJL model. The variant of the NJL model examined here, as opposed to the quark bag model, does not permit the existence of such exotic stellar objects as strange stars.

## 3. The hadron phase

The Baym-Bethe-Pethick (BBP) equation of state [24] was used to describe the hadron phase for densities corresponding to the outer and inner crust of the star. In the nuclear and supranuclear density region ($n \geq 0.1$ fm$^{-3}$) we used the relativistic mean field (RMF) model in which nucleons interact through exchange of mesons. It was assumed that the hadron phase consists of protons, neutrons, and electrons, and that the exchange particles are mesons with different transformation properties in isotope and ordinary space: isoscalar-scalar σ-mesons, isoscalar-vector ω-mesons, isovector-scalar δ-mesons, and isovector-vector ρ-mesons. In terms of the relativistic mean field model, the relativistic nonlinear lagrangian density for the hadron phase is given by

$$\begin{aligned}\mathcal{L}_{RMF} = \overline{\psi}_N \Bigg[ &\gamma^\mu \left( i\partial_\mu - g_\omega \omega_\mu(x) - \frac{1}{2} g_\rho \vec{\tau}_N \cdot \vec{\rho}_\mu(x) \right) - \\ &- \left( m_N - g_\sigma \sigma(x) - g_\delta \vec{\tau}_N \cdot \vec{\delta}(x) \right) \Bigg] \psi_N + \frac{1}{2}\left( \partial_\mu \sigma(x) \partial^\mu \sigma(x) - m_\sigma^2 \sigma(x)^2 \right) - \\ &- U(\sigma(x)) + \frac{1}{2} m_\omega^2 \omega^\mu(x) \omega_\mu(x) - \frac{1}{4} \Omega_{\mu\nu}(x) \Omega^{\mu\nu}(x) + \frac{1}{2}\left( \partial_\mu \vec{\delta}(x) \partial^\mu \vec{\delta}(x) - m_\delta^2 \vec{\delta}(x)^2 \right) + \\ &+ \frac{1}{2} m_\rho^2 \vec{\rho}^\mu(x) \vec{\rho}_\mu(x) - \frac{1}{4} \mathcal{R}_{\mu\nu}(x) \mathcal{R}^{\mu\nu}(x) + \overline{\psi}_e \left( i\gamma^\mu \partial_\mu - m_e \right)\psi_e ,\end{aligned} \tag{15}$$

where $x = x_\mu = (t, x, y, z)$, while $\sigma(x)$, $\omega_\mu(x)$, $\vec{\delta}(x)$, and $\vec{\rho}_\mu(x)$ are the fields of the σ, ω, δ, and ρ exchange mesons, respectively. $U(\sigma)$ is the nonlinear part of the potential of the σ-field [25] and is given by

$$U(\sigma) = \frac{b}{3}(g_\sigma \sigma)^3 + \frac{c}{4}(g_\sigma \sigma)^4, \tag{16}$$

where $m_N, m_e, m_\sigma, m_\omega, m_\delta, m_\rho$ are the free particle masses, $\psi_N = \begin{pmatrix} \psi_p \\ \psi_n \end{pmatrix}$ is the isospin doublet of the nucleon bispinor, $\psi_e$ is the electron wave function, $\vec{\tau}_N$ are the isospin $2 \times 2$ Pauli matrices, and $\Omega_{\mu\nu}$ and $\mathcal{R}_{\mu\nu}$ are the antisymmetric tensors of the vector fields $\omega_\mu(x)$ and $\vec{\rho}_\mu(x)$, given by

$$\Omega_{\mu\nu}(x) = \partial_\mu \omega_\nu(x) - \partial_\nu \omega_\mu(x), \quad \mathcal{R}_{\mu\nu}(x) = \partial_\mu \rho_\nu(x) - \partial_\nu \rho_\mu(x). \tag{17}$$

$g_\sigma, g_\omega, g_\delta$, and $g_\rho$ in Eq. (15) denote the coupling constants of a nucleon with the corresponding exchange mesons. In the RMF theory the meson fields $\sigma(x)$, $\omega_\mu(x)$, $\vec{\delta}(x)$, and $\vec{\rho}_\mu(x)$ are replaced by the effective mean fields $\langle \sigma \rangle$, $\langle \omega_\mu \rangle$, $\langle \vec{\delta} \rangle$, and $\langle \vec{\rho}_\mu \rangle$.

On changing the notation for the meson fields and coupling constants to

$$g_\sigma \langle \sigma \rangle \equiv \sigma, \quad g_\omega \langle \omega_0 \rangle \equiv \omega, \quad g_\delta \langle \delta^{(3)} \rangle \equiv \delta, \quad g_\rho \langle \rho_0^{(3)} \rangle \equiv \rho, \tag{18}$$

$$\left(\frac{g_\sigma}{m_\sigma}\right)^2 \equiv a_\sigma, \quad \left(\frac{g_\omega}{m_\omega}\right)^2 \equiv a_\omega, \quad \left(\frac{g_\delta}{m_\delta}\right)^2 \equiv a_\delta, \quad \left(\frac{g_\rho}{m_\rho}\right)^2 \equiv a_\rho \tag{19}$$

in the mean field approximation the Euler-Lagrange equations yield the following equations for the exchange meson fields:

$$\sigma = a_\sigma \left( n_{sp}(n_p, \sigma, \delta) + n_{sn}(n_n, \sigma, \delta) - b\sigma^2 - c\sigma^3 \right), \tag{20}$$

$$\omega = a_\omega (n_p + n_n), \tag{21}$$

$$\delta = a_\delta \left( n_{sp}(n_p, \sigma, \delta) - n_{sn}(n_n, \sigma, \delta) \right), \tag{22}$$

$$\rho = \frac{1}{2} a_\rho (n_p - n_n), \tag{23}$$

where

$$n_{sp}(n_p,\sigma,\delta) = \frac{1}{\pi^2} \int_0^{(3\pi^2 n_p)^{1/3}} \frac{m_p^*(\sigma,\delta)}{\sqrt{k^2 + m_p^*(\sigma,\delta)^2}} k^2 dk, \tag{24}$$

$$n_{sn}(n_n,\sigma,\delta) = \frac{1}{\pi^2} \int_0^{(3\pi^2 n_n)^{1/3}} \frac{m_n^*(\sigma,\delta)}{\sqrt{k^2 + m_n^*(\sigma,\delta)^2}} k^2 dk. \tag{25}$$

The effective masses of the proton and neutron are defined by

$$m_p^*(\sigma,\delta) = m_N - \sigma - \delta, \quad m_n^*(\sigma,\delta) = m_N - \sigma + \delta. \tag{26}$$

The $\beta$-equilibrium condition takes the form

$$\mu_n(n_n,\sigma,\omega,\delta,\rho) = \mu_p(n_p,\sigma,\omega,\delta,\rho) - \mu_e(n_e), \tag{27}$$

where $\mu_n(n_n,\sigma,\omega,\delta,\rho)$, $\mu_p(n_p,\sigma,\omega,\delta,\rho)$, and $\mu_e(n_e)$ are the chemical potentials of the proton, neutron, and electron, respectively, and are given by

$$\mu_p(n_p,\sigma,\omega,\delta,\rho) = \sqrt{(3\pi^2 n_p)^{2/3} + m_p^*(\sigma,\delta)^2} + \omega + \frac{1}{2}\rho, \tag{28}$$

$$\mu_n(n_n,\sigma,\omega,\delta,\rho) = \sqrt{(3\pi^2 n_n)^{2/3} + m_n^*(\sigma,\delta)^2} + \omega - \frac{1}{2}\rho, \tag{29}$$

$$\mu_e(n_e) = \sqrt{(3\pi^2 n_n)^{2/3} + m_e^2}. \tag{30}$$

For a plasma consisting of neutrons, protons, and electrons, the condition of local electrical neutrality takes the form

$$n_p - n_e = 0, \tag{31}$$

and the baryon concentration is given by

$$n_B = n_p + n_n. \tag{32}$$

For a given value of the baryon concentration $n_B$, the seven equations (20)-(23), (27), (31), and (32) can be solved numerically to determine the average fields of the $\sigma, \omega, \delta$, and $\rho$ exchange mesons, as well as the particle concentrations $n_p$, $n_n$, and $n_e$.

The energy density of the hadron phase ("npe" matter) is given by

$$\varepsilon_{npe} = \frac{1}{\pi^2} \int_0^{(3\pi^2 n_p)^{1/3}} \sqrt{k^2 + (m_N - \sigma - \delta)^2}\, k^2 dk + \frac{1}{\pi^2} \int_0^{(3\pi^2 n_n)^{1/3}} \sqrt{k^2 + (m_N - \sigma + \delta)^2}\, k^2 dk + \\ + \frac{1}{\pi^2} \int_0^{(3\pi^2 n_e)^{1/3}} \sqrt{k^2 + m_e^2}\, k^2 dk + \frac{b}{3}\sigma^3 + \frac{c}{4}\sigma^4 + \frac{1}{2}\left( \frac{\sigma^2}{a_\sigma} + \frac{\omega^2}{a_\omega} + \frac{\delta^2}{a_\delta} + \frac{\rho^2}{a_\rho} \right). \tag{33}$$

For the pressure of the hadron phase we have

$$P_{npe} = \frac{1}{3\pi^2} \int_0^{(3\pi^2 n_p)^{1/3}} \frac{k^4}{\sqrt{k^2 + (m_N - \sigma - \delta)^2}}\, dk + \frac{1}{3\pi^2} \int_0^{(3\pi^2 n_n)^{1/3}} \frac{k^4}{\sqrt{k^2 + (m_N - \sigma + \delta)^2}}\, dk + \\ + \frac{1}{3\pi^2} \int_0^{(3\pi^2 n_e)^{1/3}} \frac{k^4}{\sqrt{k^2 + m_e^2}}\, dk - \frac{b}{3}\sigma^3 - \frac{c}{4}\sigma^4 + \frac{1}{2}\left( -\frac{\sigma^2}{a_\sigma} + \frac{\omega^2}{a_\omega} - \frac{\delta^2}{a_\delta} + \frac{\rho^2}{a_\rho} \right). \tag{34}$$

When the thermodynamic parameters $n_B^{HM}$, $\varepsilon^{HM}$, and $P^{HM}$ of the hadron phase are known, it is possible to find the baryon chemical potential of the hadron phase as a function of pressure,

$$\mu_B^{HM}(P^{HM}) = \frac{P^{HM} + \varepsilon^{HM}(P^{HM})}{n_B^{HM}(P^{HM})}. \tag{35}$$

Following Refs. 26 and 4, in this paper we have chosen a value of $a_\delta = 2.5$ fm² for the parameter $a_\delta$, which characterizes the rate of interactions of a nucleon with $\delta$-mesons. The other five model parameters $a_\sigma, a_\omega, a_\rho$, $b$, and $c$ are determined beginning with empirically known values of such nuclear characteristics as the effective nucleon mass $m_N^* = 0.78 m_N$ ($m_N = 938.9$ MeV), saturation concentration of nuclear matter $n_0 = 0.153$ fm⁻³, the specific binding energy at saturation $f = -16.3$ MeV, the compression modulus K = 300 MeV, and the symmetry energy $E_{sym}^{(0)} = 32.5$ MeV. The model parameters of Ref. 4 yield $a_\sigma = 9.154$ fm², $a_\omega = 4.828$ fm², $a_\rho = 13.621$ fm², $b = 1.654 \cdot 10^{-2}$ fm⁻¹, and $c = 1.319 \cdot 10^{-2}$.

## 4. The hadron-quark phase transition and the parameters of hybrid stars

Studying the structure and integral parameters of quark-hadron hybrid stars requires knowledge of the thermodynamic characteristics of hadron and quark matter separately, as well as of the mechanism for mutual conversion of these two phases. At present there is still no generally accepted idea of how the quark-hadron phase transition takes place. It has been proposed in many studies of this problem [27-30] that each of the phases, separately, is electrically neutral and the phase transitions take place at constant pressure with a discontinuous change in energy density. In this case, the parameters of the phase transition are determined by a standard Maxwell construction. It has been shown [1] that a hadron-quark phase transition can also occur under a condition of global electrical neutrality, where both phases are electrically charged but the mixed hadron-quark phase is electrically neutral as a whole. In this case the phase equilibrium is determined by the Gibbs phase rule for which the energy density changes discontinuously during the transition.

If the coefficient of surface tension between the quark and hadron matter is less than some critical value, then the phase transition takes place with the formation of different geometric structures of mixed quark-hadron matter. For larger coefficients of surface tension, an ordinary first order phase transition is energetically favorable and its characteristics are determined by a standard Maxwell construction. That the coefficient of surface tension between quark and hadron matter is currently unknown means that it is not possible to establish which of the two above-mentioned deconfinement mechanisms takes place in the interiors of neutron stars.

In this paper we assume that surface tension effects are so strong that the quark deconfinement phase transition proceeds via a Maxwell construction scenario. The equilibrium of the two phases in this case obeys the condition

$$\mu_B^{HM}(P_0) = \mu_B^{QM}(P_0). \tag{36}$$

Table 1 lists the results of a numerical calculation of the parameters of the first order phase transition. Here $P_0$ is the pressure at which the two phases coexist and the pair of values of $n_B$ and e correspond to the concentration and energy density of the two phases at equilibrium. We note that in the NJL model examined here, the phase transition between the hadron and quark matter takes place at such high concentrations $n_B^H = 0.584$ fm$^{-3}$ and $n_B^Q = 0.814$ fm$^{-3}$ that the threshold concentration $n_B^Q$ for the quark phase is only slightly below the concentration at which a hybrid star loses stability.

TABLE 1. Parameters of the First Order Phase Transition Determined by a Maxwell Construction

|  | $P_0$, MeV/fm$^3$ | $n_B$, fm$^{-3}$ | $\varepsilon$, MeV/fm$^3$ |
|---|---|---|---|
| Hadron matter | 150.5 | $n_B^H = 0.584$ | $\varepsilon^H = 647.1$ |
| Quark matter | 150.5 | $n_B^Q = 0.814$ | $\varepsilon^Q = 960.9$ |

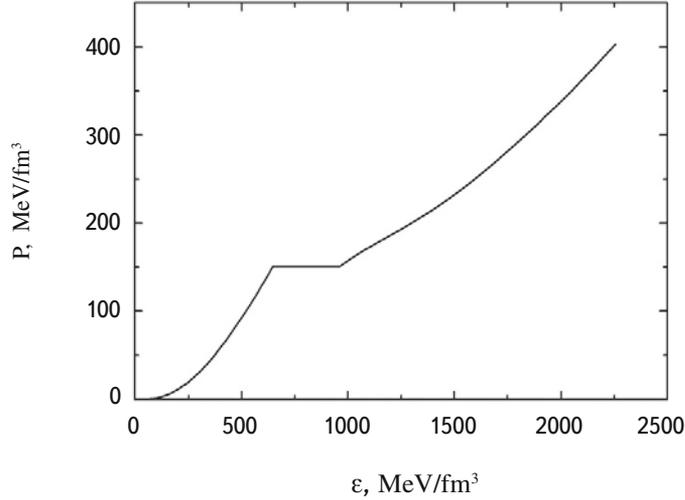

Fig. 2. The equation of state of matter for a neutron star with a hadron-quark phase transition.

Figure 2 shows the equation of state that we found for matter in a neutron star with a hadron-quark phase transition that proceeds in accord with a Maxwell scenario. Here the density discontinuity parameter $\lambda = \varepsilon_Q/(\varepsilon_H + P_0)$, which is decisive for the stability of an infinitely small quark core of a neutron star, is equal to $\lambda = 1.2$.

Since the condition $\lambda < 3/2$ holds, a hybrid star with a central core of quark matter of arbitrarily low mass will be stable in terms of this model [31].

With this equation of state and by numerical integration of the Tolman-Oppenheimer-Volkoff (TOV) equations [32,33] supplemented by equations for the moment of inertia of the star [34], we have determined various characteristics of a hybrid star for different values of the central pressure, including: gravitational mass $M$, radius $R$, quark core mass $M_{core}$ and radius $R_{core}$, moment of inertia I, and gravitational red shifts $Z_c$ for photons from the star's center and $Z_s$ from its surface.

Table 3 shows the results of our calculations. The left frame shows the dependence of the gravitational mass on the central pressure, and the right, the relationship between the mass and radius of a star. The dashed curves correspond to the case where there is no quark deconfinement phase transition. The horizontal lines correspond to the measured masses of the millisecond radio pulsars PSR J1614-2230 ($1.97 \pm 0.04\, M_\odot$) [20] and PSR J0348+0432 ($2.01 \pm 0.04\, M_\odot$) [21]. As Fig. 3 shows, the combination of models for hadron matter and quark matter examined here yields a result for the maximum gravitational mass of a hybrid star that is consistent with observational data [20,21].

Table 2 lists the major parameters of the stars in several configurations. Configuration "A" is a star with mass $1.44\, M_\odot$ corresponding to the well known Hulse-Taylor pulsar PSR B1913+16. "B" is a critical configuration with a central pressure $P_0$ at which the two phases coexist. This configuration corresponds to a star with a hadronic composition and maximum mass. At the same time, this is the configuration corresponding to a hybrid star with minimum mass. Configuration "C" corresponds to a hybrid star with maximum mass. The data in the table imply that stable hybrid stars ($dM/dP_c \geq 0$) with the variant of the equation of state examined here occupy a fairly narrow range of central densities, $\rho_c \in (1.71 \div 1.73] 10^{15}$ g/cm³. This is a direct consequence of the fact that the lower threshold for the quark phase transition is quite high in the NJL model compared to the corresponding value in the MIT bag model.

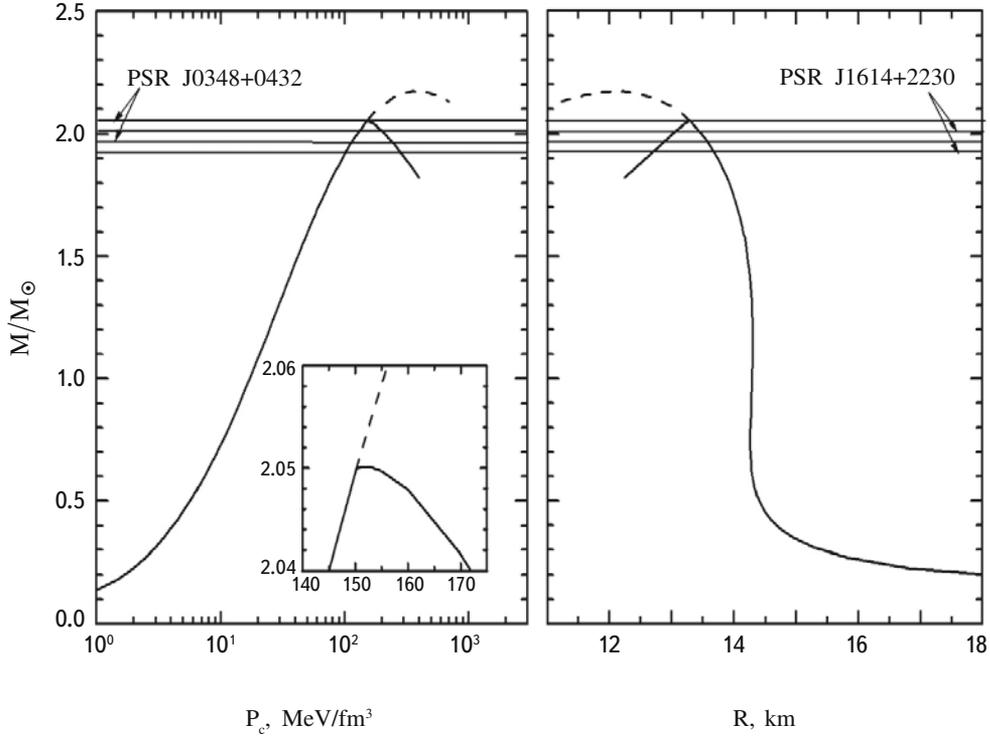

Fig. 3. Left frame: gravitational mass as a function of central pressure for a neutron star with a quark deconfinement phase transition. The inset is a fragment of the curve near the phase transition point. Right frame: the mass-radius relation for neutron stars with a phase transition. The dashed curves correspond to the case with no hadron-quark phase transition. The horizontal lines correspond to the measured masses of the millisecond radio pulsars PSR J1614-2230 ($1.97 \pm 0.04\,M_\odot$) and PSR J0348+0432 ($2.01 \pm 0.04\,M_\odot$).

TABLE 2. Major Integral Parameters for the Three Star Configurations ("A," "B," and "C")

| Configurations | $P_c$ MeV/fm$^3$ | $\rho_c$ $10^{15}$ g/cm$^3$ | $R$ km | $R_{core}$ km | $M$ $M_\odot$ | $M_{core}$ $10^{-4} M_\odot$ | $I$ $10^{45}$ gcm$^3$ | $Z_s$ | $Z_c$ |
|---|---|---|---|---|---|---|---|---|---|
| A | 38 | 0.586 | 14.25 | 0 | 1.44 | 0 | 2 | 0.194 | 0.444 |
| B | 150.5 | 1.71 | 13.313 | 0 | 2.0506 | 0 | 2.860 | 0.354 | 0.996 |
| C | 152 | 1.73 | 13.309 | 0.64 | 2.0509 | 9.56 | 2.858 | 0.355 | 0.998 |

The maximum mass $2.05 M_\odot$ of a hybrid star obtained in this paper is consistent with recent observations of radio pulsars [20,21] with masses on the order of twice the solar mass.

## 5. Conclusion

Phase transitions with deconfinement of quarks in neutron stars have been studied in this paper. An equation of state was constructed by combining an equation of state for hadron matter obtained with the relativistic mean field theory with an equation of state for quark matter obtained with a local version of the NJL model. The thermodynamic characteristics of quark matter have been calculated in terms of an NJL model and it has been shown that Witten's hypothesis that the ground state of this material at zero pressure is a β-equilibrium of three-flavor *uds* matter does not hold in this case. As opposed to the quark bag model, the version of the NJL model examined here, does not allow the existence of such exotic stellar objects as strange stars.

The parameters of the phase transition have been calculated assuming that the phase transition proceeds in accordance with a Maxwell construction and it has been shown that in the variant of the model considered here, the phase transition between hadron and quark matter takes place at fairly high concentrations $n_B^H = 0.584$ fm⁻³ and $n_B^Q = 0.814$ fm⁻³.

Integral characteristics of compact stars, such as the gravitational mass $M$, radius $R$, quark core mass $M_{core}$ and radius $R_{core}$, moment of inertia I, and gravitational red shifts $Z_c$ for photons from the star's center and $Z_s$ from its surface, have been calculated for different values of the central pressure. It has been shown that, for the equation of state examined here, stable hybrid stars correspond to a fairly narrow range of central densities $\rho_c \in (1.71 \div 1.73] 10^{15}$ g/cm³. In our model, hybrid stars are found on the same branch as neutron stars, so that a branch for a third family does not develop. The maximum gravitational mass of a stable hybrid star, $M_{max} = 2.05 M_\odot$, is attained for a central density of $\rho_m = 1.73 \cdot 10^{15}$ g/cm³. Our result for the maximum mass of a hybrid star satisfies the criterion $M_{max} > 2 M_\odot$, in agreement with observations that indicate the existence of neutron stars with masses on the order of twice the sun's. The configuration with a maximum mass in our case has a quark core with a mass on the order of $M_{core} \approx 10^{-3} M_\odot$ and a radius on the order of $R_{core} \approx 0.6$ km.

This work was done at the Scientific Research Laboratory for the Physics of Superdense Stars in the Department of the Theory of Wave Processes and Physics at Erevan State University and was supported by the State